\documentclass[twocolumn]{aastex701}

\usepackage{amsmath}
\usepackage{bm}

\newcommand{\funits}[1]{erg cm$^{-2}$ s$^{-1}$ \AA{}$^{-1}$}
\newcommand\xmm{\emph{XMM-Newton}}

\received{September 15, 2025}
\revised{October 21, 2025}
\accepted{December 2, 2025}
\submitjournal{ApJ}

\shorttitle{Campaign on AU Mic~\textsc{IV}: Radio Quiescence}
\shortauthors{Tristan et al.}

\begin{document}
\title{A 7 Day Multiwavelength Flare Campaign on AU Mic.~\textsc{IV}: \\
Quiescent Gyrosynchrotron and Gyroresonance Radiation from 12 to 25 GHz}

\author[0000-0001-5974-4758]{Isaiah I. Tristan}
\affiliation{Rice Space Institute, Rice University, Houston, TX 77005, USA}
\affiliation{Laboratory for Atmospheric and Space Physics, Boulder, CO 80303, USA}
\email[show]{isaiah.tristan@rice.edu} 

\author[0000-0001-5643-8421]{Rachel A. Osten}
\affiliation{Space Telescope Science Institute, Baltimore, MD 21218, USA}
\affiliation{Center for Astrophysical Sciences, Johns Hopkins University, Baltimore, MD 21218, USA}
\email{osten@stsci.edu}

\author[0000-0002-0412-0849]{Yuta Notsu}
\affiliation{Department of Astrophysical and Planetary Sciences, University of Colorado Boulder, CO 80305, USA}
\affiliation{Laboratory for Atmospheric and Space Physics, Boulder, CO 80303, USA}
\affiliation{National Solar Observatory, Boulder, CO 80303, USA}
\email{Yuta.Notsu@colorado.edu}

\author[0000-0001-7458-1176]{Adam F. Kowalski}
\affiliation{Department of Astrophysical and Planetary Sciences, University of Colorado Boulder, CO 80305, USA}
\affiliation{Laboratory for Atmospheric and Space Physics, Boulder, CO 80303, USA}
\affiliation{National Solar Observatory, Boulder, CO 80303, USA}
\email{adam.f.kowalski@colorado.edu}

\author[0000-0002-3699-3134]{Steven R. Cranmer}
\affiliation{Department of Astrophysical and Planetary Sciences, University of Colorado Boulder, CO 80305, USA}
\affiliation{Laboratory for Atmospheric and Space Physics, Boulder, CO 80303, USA}
\email{Steven.Cranmer@lasp.colorado.edu}

\begin{abstract}
We present an analysis of the radio quiescent data from a multiwavelength campaign of the active M-dwarf flare star AU Mic (dM1e) that occurred in October 2018. Using Ku-band data (12 to 18 GHz) from the Very Large Array and K-band data (17 to 25 GHz) from the Australia Telescope Compact Array, we find that the quiescent spectrum can be decomposed into two components: {one falling} with frequency and one that remains flat. The flat component has a relatively steady flux density of 0.64 $\pm$ 0.14 mJy. The falling component varies in strength, but exhibits a spectral index of $\alpha$ = $-0.88 \pm 0.10$. 
The falling component is thus consistent with nonthermal, optically thin gyrosynchrotron radiation with a corresponding power-law index similar to flares from AU Mic. While a flat component may arise from thermal, optically thin free-free emission, the observed flux density and inferred mass-loss rate are both too large compared to previous {stellar wind and X-ray emission} theory and models, necessitating an alternative explanation. This flat component instead matches well with an optically thick gyroresonance component integrated over multiple source regions such that the composite spectra {are} reasonably flat. The persistence of these components across the rotational period suggests multiple source regions, which may help explain changes in flux density and persistent high-energy electrons.
\end{abstract}


\section{Introduction}

M dwarfs are currently considered high priority targets in the search for habitable planets and life outside of our solar system due to their abundance in the Solar Neighborhood \citep{Henry2024}. However, much is still unknown about the differences of radiative processes between M dwarfs and solar-type stars \citep[see][for a review]{Kowalski2024c}. M-dwarf radio (i.e., cm-wavelength) observations have a wealth of past studies and theories \citep[e.g.,][]{Gudel1993, Gudel1994, White1994, Lim1996, Leto2000, Smith2005, Osten2005, Osten2006}. However, there have been few studies that examine these radio relations using current-generation technology, which offers greater sensitivity, high-cadence observations, and expanded simultaneous frequency coverage. The radio spectrum of M dwarfs above 10 GHz remains relatively unexplored, particularly regarding time-resolved variability. Strong stellar radio emission originates from nonthermal electrons accelerated by magnetic processes, providing information on magnetic field strengths, topology, and particle acceleration mechanisms in the source regions. While we have a general idea of the spectral shape from solar flares \citep{Nita2004}, differences in magnetic and atmospheric properties compared to the Sun {\citep[e.g.,][]{Allred2015} likely cause departures from these constraints in stellar flares \citep[e.g.,][]{Tristan2025}. It follows that the quiescent radio spectrum is also fairly unconstrained in stellar emissions.} 
Further, radio variability uniquely allows us to study any cycles in long-lived nonthermal particle populations or magnetic structures. Such observations will help determine M-dwarf habitability by constraining stellar wind and high-energy particle conditions, which can describe some of the space-weather impacts exoplanets experience \citep[see][and references within]{Vidotto2013}.

From 10 to 30 GHz, the spectral shape and intensity is dominated by contributions from nonthermal gyrosynchrotron emission, gyroresonance emission, and thermal free-free emission \citep{Dulk1985, Osten2006}. Nonthermal gyrosynchrotron emission originates from the acceleration of mildly relativistic particles in a strong magnetic field, and this radiation from stellar flares typically affects the 1 to 100 GHz range. During solar flares, the spectral peak frequency ($\nu_{\text{peak}}$) is typically around 4 to 6 GHz \citep{Nita2002}, though there is evidence that $\nu_{\text{peak}}$ is often, but not always, higher in M-dwarf flares \citep{Tristan2025}. Additionally, the source of radio emission above 100 GHz is still under investigation with the most likely explanations being gyrosynchrotron or relativistic synchrotron radiation \citep{MacGregor2020}.
Gyroresonance emission arises from thermal electrons in the hot plasma of the corona emitting radiation at the low harmonics of the cyclotron frequency. This emission is generally dominant at higher frequencies ($ \nu \geq 20$ GHz), occasionally resulting in U-shaped radio spectra \citep{Gudel1989}.
Free-free emission arises from electrons being deflected by ions in the ambient plasma. It is present to some degree at all frequencies. The thermal free-free emission remains optically thick at or above 34.5 GHz in solar-type stars \citep{Villadsen2014}{, and this may also be the case for cooler stars. Note that optically thin thermal free-free emission is instead expected to exhibit a flat spectrum. Multiple stellar studies have found a flat, though non-simultaneous, spectrum in the $\sim$1--40 GHz range \citep{Bastian2018, Plant2024}, but contributions from other emission mechanisms are not ruled out in these cases.
Thus,} the degree to which each of these mechanisms is present in M dwarfs, and their effect on spectral shape and evolution, remains relatively unconstrained.

To address these concerns, we analyze quiescent data from 20 hours of Ku-band (12--18 GHz) observations from the Karl G.~Jansky Very Large Array \citep[VLA;][]{Perley2011} and 40 hours of K-band (17--25 GHz) observations from the Australia Telescope Compact Array \citep[ATCA;][]{Wilson2011} collected over 5 days during an October 2018 multiwavelength campaign on AU Mic, an active, flaring dM1e star with exoplanets and a debris disk \citep[see][]{Augereau2006, Plavchan2020}. Previous campaign papers, T23 \citep{Tristan2023}, T25 \citep{Tristan2025}, and N25 \citep{Notsu2025}, provide more information and references on this target star and focus on various aspects including multiwavelength flare energetics and X-ray stellar characterization.

T25, which focuses on the energetics of radio flares that occurred during the campaign, reports a variable quiescent flux density, with the VLA Ku band between 1.2 and 3 mJy and ATCA K band between 0.8 and 2.3 mJy.  During flares studied with 10-second integrations, the $\nu_{\text{peak}}$ is often found within or above the Ku band, which may be due to the overall higher magnetic activity of M-dwarfs \citep[e.g., average magnetic field strength of 2.3 kG for AU Mic from Table 1 of][]{Kochukhov2021} compared with the Sun \citep[about 1 G;][]{Babcock1955}. The Ku-band frequency range is thus of interest, as analyzing both the optically thick region below and the optically thin regions above the $\nu_{\text{peak}}$ are necessary to fully characterize the radio emitting source.
However, the quiescent signal in short integrations was ambiguous due to {the scatter of the light curve per frequency bin}. A high $\nu_{\text{peak}}$ in quiescence could indicate {similarities} in electron energies, magnetic field strengths, or loop structures between persistent stellar source regions and {the temporary conditions of solar flares following magnetic reconnection} \citep[see][and discussion in T25]{Dulk1985}. Characterizing the quiescent signal is key to understanding magnetic cycles and the conditions under which flares are produced. 
Here, we follow up using longer integrations to characterize variability and bulk properties, improving our knowledge of M-dwarf quiescent radio activity.

In this work, we present the first in-depth analysis of the high-frequency radio spectrum of AU Mic that also uses simultaneous, multiwavelength observations to characterize emission mechanisms.
We aim to highlight the importance of quiescent and low-time-resolution radio data for complementing high-time-resolution flare studies and understanding magnetic activity of M-dwarf stars.

\section{Data Summary}
\label{sec:data_summary}

Data collection, calibration, and reduction are detailed in \S{2} and Appendix A of T25. Briefly, there are 13.62 and 39 hours of on-source time with the VLA Ku band and ATCA K band, respectively. VLA data are reduced using the \texttt{CASA} software \citep{CASA_2022}, version 6.6.4, while ATCA data are reduced using \texttt{MIRIAD} \citep{Sault1995}. The \texttt{CASA} task \texttt{uvmodelfit} is used to fit visibility data in time bins, calculating flux densities and errors separately for each correlation (\emph{RR} and \emph{LL} for VLA, \emph{XX} and \emph{YY} for ATCA). These values are consistent with baseline averaging, which is used to calculate the \emph{XY} and \emph{YX} ATCA terms.

The Stokes parameters for total intensity ($I$) and circular polarization ($V$) are calculated as 
\begin{align}
    \text{VLA:}\ \ I &= \frac{RR+LL}{2},\ V = \frac{RR-LL}{2}, \\
    \text{ATCA:}\ \ I &= \frac{XX+YY}{2},\ V = \frac{XY-YX}{2i},
\end{align}
with uncertainties propagated accordingly \citep[see][]{Osten2004}. Circular polarization fraction is computed as $\pi_c(\%)=V/I \times 100$. To create radio spectra, the subbands are split evenly into quadrants and analyzed similarly to the total light curves.

\begin{figure*}[!ht]
\centering
\includegraphics[width=1\linewidth]{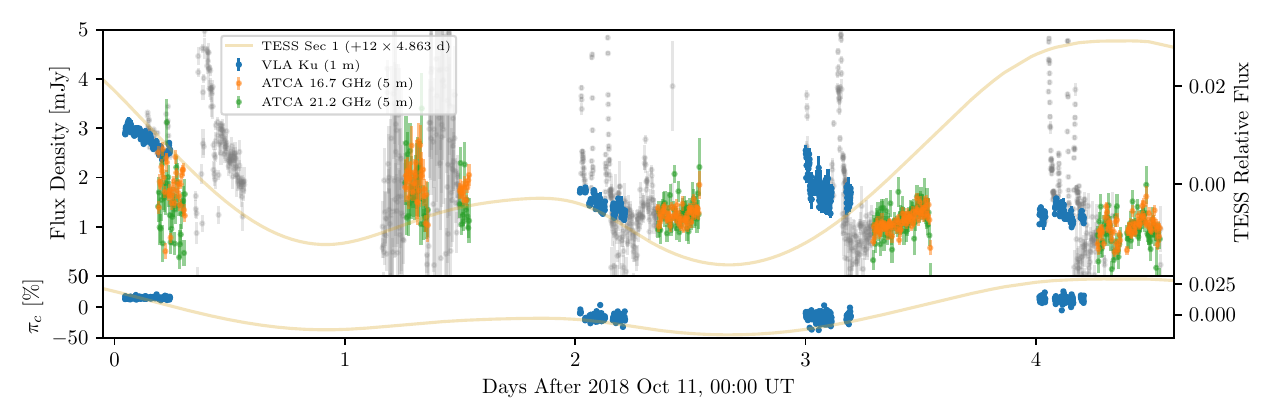}
\caption{\textit{Top:} Light curves for the VLA and ATCA are shown, in bins of {1 minute (1 m) and 5 minutes (5 m)}, respectively. Flares and periods of high noise due to weather effects are grayed out and not used in calculations.
The rotational modulation extrapolated from the temporally nearest TESS Sector is shown in yellow {\citep[cf., Figure 1(a) of][]{Ikuta2023}}.
\textit{Bottom:} Circular polarization fraction of the VLA data is displayed against the TESS rotational modulation. The gray dashed line marks $\pi_c = 0$.
\label{fig:TESSRotation}
}
\end{figure*}

Light curves for the 1-minute VLA wideband and 5-minute ATCA 16.7 and 21.2 GHz bands are shown in Figure \ref{fig:TESSRotation} and compared against the white-light rotational modulation curve from the Transiting Exoplanet Survey Satellite \citep[TESS;][]{Ricker2015}. Nearly all Stokes I data here achieve signal-to-noise ratios greater than 3 (SNR $>$ 3), though no significant quiescent Stokes V detections are found in the ATCA data.

\section{Quiescent Analysis}
\label{sec:quiscent_analysis}

\subsection{Spectral Indices Per Observing Period}
\label{sec:spectral_index_per_day}
\begin{figure}[!ht]
\centering
\includegraphics[width=1\linewidth]{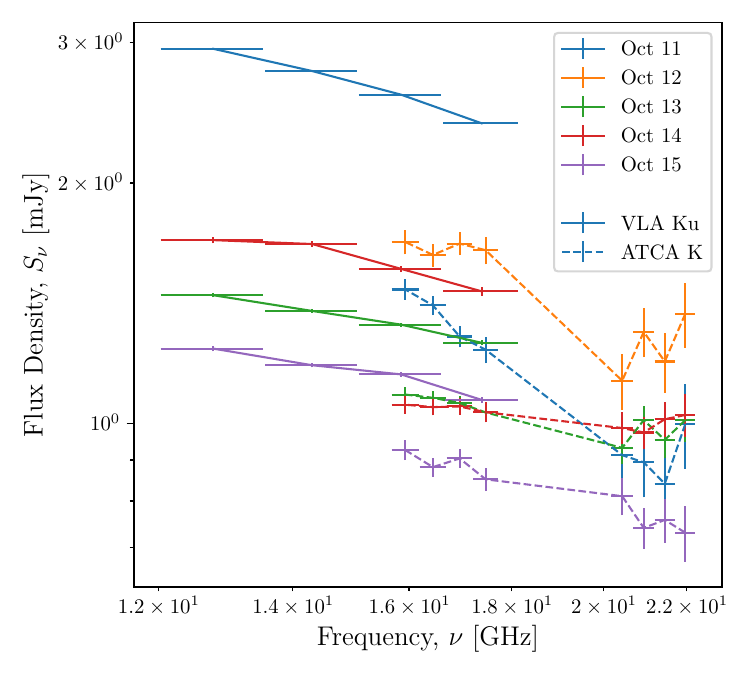}
\caption{Radio spectra integrated over all quiescent times.
The VLA Ku-band wideband and the 2 ATCA K-band subbands are split evenly between 4 frequency ranges each.
\label{fig:FigQuiescentAll}
}
\end{figure}

In T25, VLA 10-second-binned quiescent spectra display a mix of falling, flat, and rising spectra in the Ku band. To investigate this, we {fit visibility data from} all quiescent times per day to form {integrated} spectra for the VLA and ATCA (Figure \ref{fig:FigQuiescentAll}). There are clear variations per day in both slope and intensity, and the decrease in flux density with time is not monotonic. To quantify the slope, we measure the spectral index, defined as 
\begin{equation}
    \alpha = \frac{\log_{10}(S_{\nu,2}/S_{\nu,1})}{\log_{10}(\nu_2/\nu_1)}, \label{eq:specind}
\end{equation}
where $S_\nu$ is flux density and $\nu$ is the representative frequency where $\nu_1 < \nu_2$ (see Table \ref{tbl:quiescent_levels}). In optically thin gyrosynchrotron radiation, this is directly related to the power-law index of the emitting particle distribution ($\delta_r$). From solar estimates, common values are between $-0.58$ and $-5.08$, where $\alpha = 1.22 - 0.9\delta_r$ for $2<\delta_r<7$ \citep{Dulk1985}. 

\begin{deluxetable*}{cccccc}
\tablewidth{0pt}
\tablecaption{Characteristic Radio Quiescent \label{tbl:quiescent_levels}}
\tablehead{
\colhead{Inst.} & \colhead{Date} & \colhead{$\alpha_{\text{tot}}$} & \colhead{Stokes $I$} & \colhead{Stokes $V$} & \colhead{$\sigma_I$}  \\ 
 & & & (mJy) & (mJy) & (mJy) 
}
\startdata 
VLA        & Oct 11 & $-0.68 \pm 0.05$ & $2.88 \pm 0.04$ & $+0.38 \pm 0.04$ & $0.20$ \\
12--18 GHz & Oct 13 & $-0.42 \pm 0.03$ & $1.12 \pm 0.04$ & $-0.12 \pm 0.04$ & $0.16$ \\
           & Oct 14 & $-0.46 \pm 0.13$ & $1.62 \pm 0.09$ & $-0.38 \pm 0.09$ & $0.26$ \\
           & Oct 15 & $-0.39 \pm 0.06$ & $1.12 \pm 0.04$ & $+0.12 \pm 0.04$ & $0.13$ \\
\hline
    ATCA       & Oct 11 & $-1.92 \pm 0.09$ & $1.62 \pm 0.11$, $1.62 \pm 0.31$ & \nodata & $0.45$, $0.45$ \\
16.7, 21.2 GHz & Oct 12 & $-0.80 \pm 0.21$ & $1.88 \pm 0.11$, $1.12 \pm 0.21$ & \nodata & $0.35$, $0.55$ \\
               & Oct 13 & $-0.56 \pm 0.05$ & $1.12 \pm 0.09$, $1.12 \pm 0.19$ & \nodata & $0.16$, $0.45$ \\
               & Oct 14 & $-0.20 \pm 0.05$ & $1.12 \pm 0.09$, $1.12 \pm 0.19$ & \nodata & $0.20$, $0.23$ \\
               & Oct 15 & $-0.68 \pm 0.09$ & $1.12 \pm 0.09$, $0.88 \pm 0.19$ & \nodata & $0.23$, $0.25$ \\
\enddata
\tablecomments{
$\alpha_{\text{tot}}$ is the spectral index calculated over the total integrated spectra from Figure \ref{fig:FigQuiescentAll} {before modeling for separate components}, and $\sigma_I$ is the standard deviation of the quiescent light curve. Stokes values are the modes of the quiescent values and errors in the 1-minute VLA and 5-minute ATCA data (see Section \ref{sec:lowtimeres}). Only $\sim$50\% of the October 14--15 VLA Stokes \textit{V} data achieve an SNR $>$ 3 and are included.
}
\end{deluxetable*}

Of the VLA data, the spectral index of October 11 at $\alpha = -0.68$ is consistent with optically thin gyrosynchrotron radiation with a power-law index of $\delta \approx 2.1$.
However, there are no flares during, or even temporally close to, these observations in multiwavelength X-ray Multi-Mirror Mission (\xmm{}) OM UVW2 and EPIC-pn X-ray data (see Figure 2(a) of T25). This may indicate that this gyrosynchrotron emission is caused by ambient trapped electrons in the magnetic fields of AU Mic, rather than from a gyrosynchrotron flare event. Note that October 14 has a flatter edge in the lower frequencies. This may indicate high $\nu_{\text{peak}}$ values near 12 GHz, though wider simultaneous coverage would be needed to confirm this. 

The other days exhibit flatter spectra, indicating that they may result from a combination of emission types.
{Note, a flatter gyrosynchrotron spectrum is possible if a high-energy cutoff is imposed and a significant fraction of particles have higher-end energies. The original cutoff of $\delta_r = 2$ from the \citet{Dulk1985} approximations likely comes from diffusive shock acceleration \citep[e.g.,][]{Bell1978} as a method of accelerating particles in a magnetized plasma, as even ultra-relativistic shocks result in $\delta > 2$ \citep{Achterberg2001} under this theory.
Magnetic reconnection events accelerating nonthermal particles, while still poorly understood, may break this perceived limit based on more recent simulations \citep{Guo2014}.
However, if the observed radiation originates from similar magnetic-reconnection processes across days, high-energy electrons dominating should also result in increased emissions at higher frequencies, which is not observed here. 
Observing near $\nu_{\text{peak}}$ can also result in {smaller $|\alpha|$ values apparently implying} $\delta_r < 2$. However, $\nu_{\text{peak}}$ can change based on source region properties like magnetic field strength and inhomogeneous structure \citep{Dulk1985}, rather than solely relying on the electron power-law distribution.
Thus, observing multiple sources over the inhomogeneous surface of AU Mic is the most likely explanation.}

To investigate the different contributions in the VLA data, we perform non-linear least-squares minimization and curve-fitting to the spectra using one-, two-, and three-component models,
\begin{align}
    S_{\nu} &= A \nu^\alpha, \\
    S_{\nu} &= A \nu^\alpha + C, \text{ and}\\
    S_{\nu} &= A \nu^\alpha + B \nu^\beta + C.
\end{align}
where $\alpha$ is the power-law slope of the assumed optically thin gyrosynchrotron component, $\beta$ is the power-law slope of the assumed optically thick gyroresonance component, and $C$ is a flat-component from thermal free-free radiation. We set {ranges of $A,\ B,\ C \geq 0$,} $-5 \leq \alpha \leq 0$, and $0 \leq \beta \leq 5$, {and vary initial guesses to search for converging solutions.} Last, we assume that the source regions are similar in composition such that $\alpha$, $\beta$, and $C$ are constant between days.

We find fits of $\alpha = -0.58 \pm 0.04$ for the one-component model and $\alpha = -0.88 \pm 0.10$ and $C = 0.64 \pm 0.14$ mJy for the two-component model. {To further ensure stable solutions, we set the initial guess values of $\alpha$ and $C$ to these and vary them by $\pm1$, which does not result in any changes.
While the three-component model does converge at solutions near $B=0$, the associated errors are too large too be statistically significant (i.e., $>100\%$). Holding $\beta=2$ for optically thick gyroresonance of a homogeneous source region does not improve this.
}

To test if either model is preferred, we employ a Bayesian Information Criterion \citep[BIC;][]{Burnham2004}, 
\begin{equation}
    \text{BIC} = k \ln(n) - 2\ln(\hat{L})
\end{equation}
where $k$ is the number of free parameters in the model, $n$ is the number of data points used in the fitting, and $\hat{L}$ is the maximum likelihood of the model. Assuming errors are distributed normally,
\begin{equation}
    \hat{L} = \frac{RSS}{n},
\end{equation}
where $RSS$ is the residual sum of squares of the fit. A lower BIC indicates a more preferred model, with a {difference} between 5 and 9 indicating a preference and a change above 10 indicating a strongly favored model.
The BIC of the two-component model is lower by 8 points, indicating that it is a better fit despite the added complexity. This model is also more consistent with a physical interpretation that the radio signal is from a combination of source regions and radiation types.

\subsection{{ATCA K-band Considerations}}
The ATCA K-band quiescent {spectral fluxes} per day are much lower than their VLA counterparts. The stark differences in Figure \ref{fig:FigQuiescentAll} may belong to instrumental, weather, and reduction factors rather than physical variability. If a calibrator is somewhat polarized when employing linear feeds, Stokes \emph{I} values will be underestimated (see Section 11 of the \texttt{MIRIAD} manual\footnote{\url{https://lweb.cfa.harvard.edu/sma/miriad/manuals/ATNFuserguide_US.pdf}}). Higher-frequency, linear-feed observations are also more likely to have adverse effects at lower elevations. This can be seen from examining Figure \ref{fig:TESSRotation}, as the start of observations exhibit higher uncertainties and lower flux densities, especially during October 13 to 15 when the gyrosynchrotron component is weaker. Both of these issues decline as AU Mic rises in elevation.
Humidity effects in the ATCA K band are also non-negligible, as a major water line lies within these frequencies (see Section 2 of T25). The most prominent case of this occurs during October 12, when the light curve is dominated by humidity-related noise. We use the low-elevation data during October 11 and less-humid data during October 12, as there are no better quiescent times available. The ATCA October 14 spectrum exhibits a larger decrease from the VLA flux density, matching the sharper decline of the VLA October 14 light curve in Figure \ref{fig:TESSRotation}. 

Despite these issues, these spectra may come from the same populations per day {since the decreases} in flux density are similar. We perform the same two-component test {for the ATCA data} from 16 to 18 GHz for October 11 and 15 {(i.e.\ overlapping times and frequencies with the VLA Ku band)}. {We find two converging solutions\footnote{The two ATCA solutions converge at $C=0.74\pm0.10$ mJy, $\alpha = -4.4\pm1.0$ and $C \approx 0$ mJy, $\alpha = -0.7 \pm 3.0$. However, the errors for $A$ in the first case and all errors in the second case are too large to be significant.}, 
both of which are not statistically significant due to high errors.}
{Note that} October 13 and 14 {exhibit flatter spectra than October 15} and do not {result in well-fitted solutions when included in} the two-component test. 
{These issues could be due to} changes in the source populations or data concerns {(e.g., using the noisy data of October 11)}.

The only indications of rising spectra occur in ATCA data, specifically the total 21.2 GHz integrations during October 11 and 12. Both of these are within error and {show falling spectra} between the 16.7 and 21.2 GHz subbands, so it is difficult to attribute any potential rising spectra to optically thick contributions from a homogeneous gyroresonance or gyrosynchrotron source. The lack of viable fits during the least-squares minimization further corroborates that any rising spectra within this frequency range are not significant.

\subsection{Origin of the Flat Component}
In the simplest case, a flat component may correspond to optically thin, thermal free-free radiation. However, we cannot easily conclude this from stellar observations. Optically thick thermal free-free emission exhibits {a rising spectrum}, but the {average} turnover frequency is currently unknown in M-dwarfs. {Note that the turnover frequency for any one radio source depends on the temperature and electron density, so there may be multiple contributions in observations of the stellar surface.} Further, simultaneous radiation by other radio-emitting processes may result in a flat composite spectrum. Here, we explore a few possibilities.

Thermal free-free emission is thought to arise from the acceleration of electrons in the ionized plasma of the stellar corona and/or wind, and the resulting flux density has a analytical form 
\begin{equation}
    S_\nu \propto  \frac{VEM}{\sqrt{T}},
\end{equation}
where $VEM$ is the volume emission measure and $T$ is the temperature, both estimated from X-ray observations \citep{Gary1981}.  \citet{Leto2000} reports a value of $S_\nu = 0.025$ mJy for $T=2.0 \times 10^7$ K and $VEM = 3.68 \times 10^{52}$ cm$^{-3}$ for AU Mic. Using average values from October 10 to 15 (see Table 1 of N25) of $T_{\text{avg}} = 9.33 \times 10^{6}$ K and $EM_{\text{tot}}=2.6 \times 10^{52}$ cm$^{-3}$, {$S_\nu \approx 0.026$ mJy. We also sum the contributions from each pair of $T$ and $VEM$ in the 10-component X-ray quiescent model in Table 1 of N25 to test differences between the average and total values. This results in $S_\nu \approx 0.032$ mJy.
These estimates are} much lower than the flat component.

\begin{figure}[!ht]
\centering
\includegraphics[width=1\linewidth]{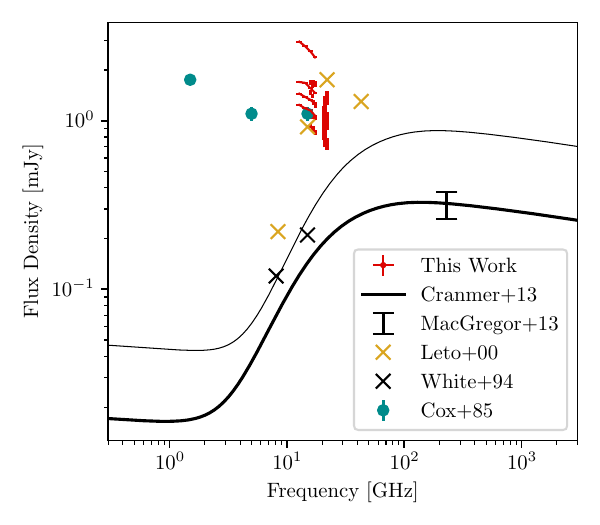}
\caption{Best-fit coronal heating model of thermal free-free emission \citep[thick solid line;][]{Cranmer2013} compared to past VLA \citep{Cox1985, White1994, Leto2000} and ALMA \citep{MacGregor2013} data. The thin solid line marks the coronal heating model limit where $\dot{M}$ rises above $10^{-8} \, M_{\odot}$~yr$^{-1}$. Spectra from Figure \ref{fig:FigQuiescentAll} are plotted in red. Cross markers indicate upper limits.
\label{fig:DataHistory}
}
\end{figure}

\citet{Cranmer2013} constructed coronal heating models that matched
both X-ray and mm measurements for AU~Mic.
These models assumed only thermal free-free emission from closed loops
(i.e., no stellar wind), and the optimal set of coronal heating
parameters produced a spectrum that is rising in the Ku-band frequencies and falls well below some of the
Ku-band observational data (see Figure \ref{fig:DataHistory}).
These models can also be passed through a cool-star wind model
\citep{Cranmer2011}, and the above model is consistent with a mass-loss
rate of $\dot{M} \approx 1.6 \times 10^{-9} \, M_{\odot}$~yr$^{-1}$. 

Given that small flares are seen each day, the stellar wind is optically thin to this radio emission in the atmosphere and corona, likely down to near the stellar surface \citep[see][]{Lim1996, Fichtinger2017}. From Equation 2 of \citet{Lim1996}, we can estimate the mass-loss rate at the optically thick limit with 
\begin{align}
    \frac{R(\nu)}{R_\sun{}} \approx 6 &\left(\frac{\nu}{10 \text{ GHz}} \frac{v_\infty}{300 \text{ km s}^{-1}}\right)^{-2/3} \left(\frac{T_{\text{wind}}}{10^4 \text{ K}}\right)^{-1/2} \nonumber \\
    \times&\left(\frac{\dot{M}}{10^{-10} \text{ M$_\sun{}$ yr}^{-1}}\right)^{2/3},
\end{align}
where $R(\nu)$ is set to the stellar surface \citep[$0.75\ R_\sun{}$ for AU Mic;][]{Plavchan2020}, $v_\infty{}$ is the terminal velocity of the wind, and $T_{\text{wind}}$ is the temperature of the wind. Using $T_{\text{wind}} = 10^4$ K and $v_\infty{} = 300$ km s$^{-1}$, $\dot{M} \approx 6.6 \times 10^{-12}$ M$_\sun{}$ yr$^{-1}$. However, hotter winds can produce higher mass-loss rates, assuming the radio emission stays optically thin {(i.e., the stellar wind
cannot overpower the observed emission from coronal sources)}. \citet{Lim1996} further states that soft X-ray estimates can provide an extreme upper limit, if the X-ray emission is assumed to come mostly from compact coronal features. Using $T_{\text{wind}} \approx 10^7$ K ($T_{avg}$ in N25) and $v_\infty{} = 600$ km s$^{-1}$ (high solar wind value), the upper limit is $\dot{M} \approx 2.4 \times 10^{-9}$ M$_\sun{}$ yr$^{-1}$, which is close to the estimated mass loss from the coronal heating models that fit previous X-ray and mm measurements. \citet{Cranmer2013} further reports a range of models that can match with the X-ray emission while producing enhanced radio emission. However, the next highest model with about 0.3 mJy at 15 GHz returns a mass-loss rate above $10^{-8}$ M$_{\odot}$~yr$^{-1}$.
Thus, we can generally rule out that the entirety of the flat component at about 0.64 mJy is due to thermal free-free emission, as the mass-loss would be far beyond the estimated physical limits.

{It is important to address that there are various methods for estimating stellar mass-loss rates.
\citet{Johnstone2015b} proposes that the mass-loss rate per unit surface area of a star is related to its mass and rotation rate by a power law. Using their Equation 4 (with $a=1.33$ and $b=-3.36$, along with $R_*/R_{\odot}=0.75$, $M_*/M_{\odot} = 0.5$, and $\Omega_*/\Omega_{\odot} = 5.15$ \citep{Plavchan2020}), $\dot{M}_{*} \approx 51~\dot{M}_{\odot}$ or $10^{-12}$ M$_{\odot}$ yr$^{-1}$. \citet{Strubbe2006} estimates the mass-loss rate by instead studying the debris disk of AU Mic and determining the stellar wind that could result in its composition. They find $\dot{M}_{*} \leq 10~\dot{M}_{\odot}$ is sufficient to explain the disk morphology, assuming the dust grains are porous. Even in the case of solid grains, $\dot{M}_* \approx 100~\dot{M}_{\odot}$. Finally, \citet{Vidotto2021} finds a rough correlation between X-ray surface flux and mass-loss rates. Using their Equation 17 with L$_X$ $\approx 2.81 \times 10^{29}$ erg s$^{-1}$ (see N25), $\dot{M}_* \approx 60~\dot{M}_{\odot}$, though the uncertainties in the equation result in values from 1 to 2000 $\dot{M}_{\odot}$.}

These mass-loss rates are lower than those of the \citet{Cranmer2013} model, which warrants a closer look.
The high model estimates arise from the photospheric velocity amplitude ($v_\perp$) outputs from the coronal heating model being used in a separate cool-star wind model. This parameter is sensitive to the X-ray luminosity (see their Figure 3). The X-ray luminosity during these observations is slightly lower than the ROSAT value used in the model at log$_{10}$L$_X$ [erg s$^{-1}$] $\approx 29.4$ when using a rough conversion with WebPIMMS \citep{Mukai1993}. This allows for smaller $v_\perp$ values while staying consistent with mm observations, leading to mass-loss rate estimates $\leq10^{-11}$  M$_{\odot}$ yr$^{-1}$. However, the $v_\perp$ needed to be consistent with the flat component again gives a mass-loss rate around 10$^{-9}$ M$_{\odot}$ yr$^{-1}$ and would not be consistent with the mm constraints. While this is closer to the extreme limits, there has been no evidence of such strong winds on AU Mic{, and multiple mass-loss rate estimates unrelated to the emission mechanism here suggest much lower values.}

One interesting result comes from treating the flat component as an optically thick gyroresonance source. We first assume that the flat and largely unpolarized spectrum is due to a superposition of spectra from different gyroresonance source regions, angles, polarizations, and harmonics. We then estimate the flux density from this optically thick source as
\begin{align}
    S_\nu \text{ [mJy]}=&~2.24 \frac{T}{2\times10^7 \text{ K}} \left( \frac{d}{5 \text{ pc}} \right)^{-2} \nonumber \\
    &\times \left(\frac{\nu}{15 \text{ GHz}}\frac{R(\nu)}{0.5 ~R_\sun{}}\right)^2
\end{align}
following \citet{Leto2000}. Using $T_{avg}$ from N25, $\nu=15$ GHz, $R(\nu)$ as the stellar surface, and $d=9.72$ pc  \citep{GaiaDR2} returns $S_\nu = 0.62$ mJy.  If $R(\nu) = 1.1~R_*$ for a compact, optically thick source \citep[as in][]{Osten2006} that would still allow for flaring emissions from higher source regions, the flux density is still within the error of the flat component. This estimate would allow for the flat component to represent both the thermal free-free emission and gyroresonance emissions from different regions of the star.

While this simplified model fits well to the three components discussed, the flat component may be more complicated in reality. However, more observations, including simultaneous observations over a wider range of frequencies, would be needed to observe long-term changes, determine a true minimum baseline, and use modeling to de-couple emissions from more sources with various observing conditions (e.g., line-of-sight angles, magnetic field strengths). The ``flat'' component here may better represent the overall magnetic activity of the star, while larger and dominant deviations like the gyrosynchrotron component come from strong, individual source regions, similar to flaring activity.

\subsection{Low-Time-Resolution Light Curves}
\label{sec:lowtimeres}

The radio quiescence exhibits variations on short timescales (see Figure \ref{fig:TESSRotation}) that are not completely captured by the longer-term integrated spectra Figure \ref{fig:FigQuiescentAll}).
To quantify the differences in average flux density between days, we bin the quiescent VLA and ATCA data separately into flux density bins of width 0.25 mJy. Associated errors are also sorted into bins of width 0.025 mJy.
The modes of these bins between the VLA and ATCA for each day are more consistent than the longer-term integrated values, as is expected given consistent measurements between observatories, and are reported in Table \ref{tbl:quiescent_levels}.

The largest difference occurs on October 11, however this can be explained by both physical and observational inferences. The VLA light curve decreases over this period, likely from a dampening of the gyrosynchrotron component, and the only quiescent times available are when AU Mic is at a lower elevation, which exhibits lower flux densities and higher noise for each observing period due to either airmass/pointing effects, daily weather patterns, or a combination of the two (refer to Section \ref{sec:spectral_index_per_day}). The standard deviations of the light curve ($\sigma_I$) per day are also calculated from these regions (as in Section 3 of T23). This method is also applied to the {10-second} VLA and {1-minute} ATCA light curves from T25, which give consistent modes with higher uncertainties.

The overall flux density levels do not vary with the stellar rotational modulation as seen in the white-light radiation, which is compared by extrapolating the temporally nearby TESS Sector 1 (2018 August) data of AU Mic. Similarly, simultaneous X-ray or H$\alpha$ data do not show modulation consistent with the radio (see Section 5.1 of N25). Further, there appear to be {occasional short-term, periodic variations with various frequencies and amplitudes}. However, these are left to a future study on quasi-period pulsations present in the radio data, including flares.

The circular polarization is at a low level ($\pi_c < 30$\%) and changes sign throughout the rotational period, which is consistent with gyrosynchrotron radiation from an extended source and further motivates our physical interpretation of the dominant component. We do not observe the part of the rotational period where the change occurs, however, nor do we have enough temporal coverage to determine if there are any signatures of auroral emission, as in the lower frequencies \citep{Bloot2024}. 

\begin{figure}[!ht]
\centering
\includegraphics[width=1\linewidth]{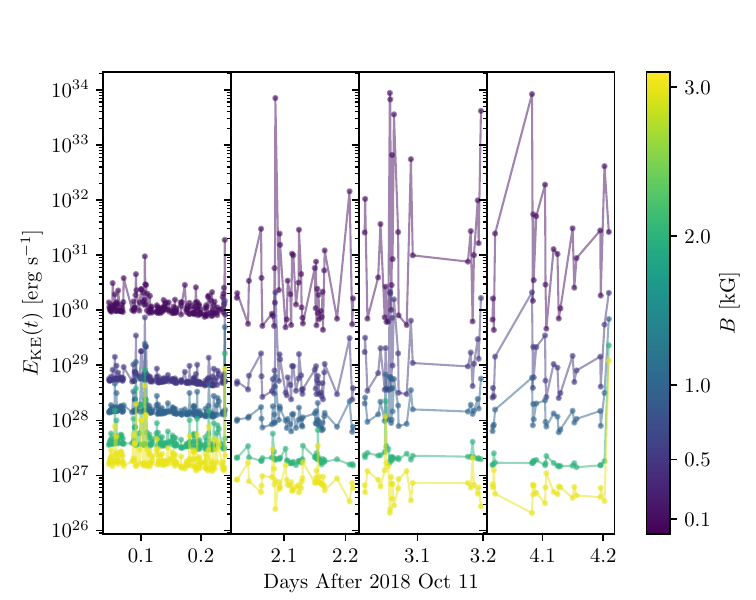}
\caption{Estimated electron kinetic energy rates ($E_\text{KE}(t)$) from the gyrosynchrotron component of the quiescent light curve. Various magnetic field strengths ($B$) are tested within a plausible range for M-dwarf atmospheres.
\label{fig:ElectronKineticEnergies}
}
\end{figure}

Given the high SNR of the VLA data, we are able to calculate the spectral index for many individual time bins. Since the spectral indices correspond to optically thin gyrosynchrotron radiation, we then calculate the power-law index of the electron distribution. By modeling the gyrosynchrotron spectrum according to the standard form from \citet{Dulk1985} with a peak frequency of 10 GHz and a low-energy cutoff of 10 keV, we can then estimate the electron kinetic energy rate ($E_\text{KE}(t)$) for a given magnetic field strength ($B$). Extensive details for this calculation are given in Section 3.5 of T25, as well as an extended discussion on critical assumptions. Here, we subtract a flat component of 0.64 mJy before calculating the spectral index. We also only use time bins with an SNR$_\alpha$ $>$ 3, which only about 30\% of data from October 13, 14, and 15 {achieved} after subtracting the flat component. Results are shown in Figure \ref{fig:ElectronKineticEnergies} for various magnetic field strengths.

If $B \sim 100$ G, electron kinetic energies from emitting sources are high enough to be comparable to flares over hour-timescales (assuming flaring energies of $10^{32}$ to $10^{34}$ from T25). Besides an unknown value for $B$, these estimates have many limitations due to assumptions in the energy calculation, like wide-band spectral shape, source orientation, and low-energy cutoff values. Future observations covering a wider range of frequencies will help determine whether these assumptions hold in M-dwarf flares, though recent radiative-hydrodynamic models are finding better agreements with much larger low-energy cutoffs \citep{Kowalski2017A, Kowalski2022, Kowalski2024b}. See Section 4.2 of T25 for more discussion on this aspect for flares.

\section{Discussion}
\label{sec:discussion}

Given the strong magnetic fields of M-dwarfs \citep[e.g.,][]{Kochukhov2021}, it may not be unusual for the quiescent spectra to be dominated by gyrosynchrotron radiation, either from many small flares \citep[for general discussion on micro-flaring on AU Mic, see][]{Robinson2001} {or} trapped particles from large flaring events.
{While power-law index values within $2 \leq \delta_r \leq 7$ are used to calculate $E_\text{KE}(t)$, $\sim$80\% of the data exhibited values within $2 \leq \delta_r \leq 3$.} Significant outliers are likely due to the scatter of the light curve within the smaller frequency subbands. This value range indicates a reasonably hard energy distribution. These values are similar to those of the optically thin flares reported in T25 and imply a common origin. 
However, $\delta_r$ is known to decrease over the course of flares \citep[e.g., Figure 2 of ][]{Osten2005}, so a relatively constant $\alpha$ despite changes in flux density may imply that the emitting energetic electrons are resupplied through frequent magnetic reconnection events \citep[see][]{Gudel1993, Gudel1994}.
The energy rates are much lower than for flares and would require low magnetic field strengths and long integrations to achieve similar total energies ($10^{32}$ to $10^{34}$ erg). Thus, the extended gyrosynchrotron radiation may also be due to trapped electrons from magnetic reconnection events that do not spark full multiwavelength flares due to a lack of impulsive energy deposition. The possibilities of frequent magnetic reconnection events and magnetic trapping are not exclusive, and a mixture would further contribute to the plausibility of constant gyrosynchrotron radiation.

The gyrosynchrotron component is present to some degree at all observed times, indicating either an extended single source region away from the equator or multiple similar source regions around the surface of the star. 
The radiation from October 11 is much stronger, which implies either larger or more magnetic reconnection events occur. If this was from a single source, we may expect changes in $\alpha$ as portions of the electrons precipitate. This not seen within $\sim$5 hours, which implies either efficient trapping over day-timescales or continuous injection. Given the decay times of flares (minutes to hours in T23/T25), it seems unlikely for this to be from a single source without a major difference in physical processes.
A persistent, low circular polarization ($\pi_c < 30$\%) is expected of stellar gyrosynchrotron sources \citep[see][]{Dulk1985}, as observed in both optically thick and thin flares in T25. The polarization reverses over the course of the rotational period, which is expected from line-of-sight effects if the source regions are focused more along one hemisphere (fitting with either a single or multiple source regions).

These quiescent flux densities of AU Mic are mostly in line with previous observations, though there have been non-detections at lower limits (see Figure \ref{fig:DataHistory}). While previous data were not as sensitive as the JVLA data presented here, this still indicates a high amount of variability, both in gyrosynchrotron and gyroresonance components. For example, the non-detections by \citet{White1994} indicate that the flat component must have also fallen below 0.2 mJy, which is within the realm of possibility given the thermal free-free emission estimation without a strong gyrosynchrotron component. Observations during similar periods of inactivity may help determine whether the thermal free-free emission is indeed optically thin (flat spectrum) at these frequencies or optically thick (rising spectrum) as predicted by coronal heating models. These observations will lead to better estimates of stellar mass-loss rates and determine appropriate values which stay within theoretical limits. However, inactive periods are likely rare on AU Mic {due} to its magnetic activity, so other stars may be better targets for a study of thermal radio processes.

{The} relatively persistent negative spectral index in this high-time-resolution data may explain why M-dwarf radio emission was previously thought to be dominated by gyrosynchrotron emission for most times, unless a strong gyroresonance emission dominated temporarily causing a U-shaped spectra \citep[e.g.,][]{Cox1985}. These temporary events have been reported for other M dwarfs {\citep[see][]{Gudel1989}}, indicating similar processes, though this was not observed with significance for any extended period here.
We also note that the emission at lower frequencies has been found to be stochastic, with Stokes I detections  $S_\nu \leq$ 4 mJy at 1 to 3 GHz \citep[based on ~250 hr of ATCA L-band monitoring;][]{Bloot2024}. However, it is currently unknown how the gyrosynchrotron component measured here affects the L-band frequencies and to what extent these variations are linked. 

\section{Conclusions}

In this work, we analyze the radio quiescence of AU Mic (dM1e) from 12 to 25 GHz, focusing on the 12 to 18 GHz range using sensitive VLA Ku-band data. Our findings are summarized as the following:
\begin{enumerate}
    \item Using integrated spectra from five consecutive days of observations, we are able to disentangle a flat, steady spectral component from a component that falls with frequency.
    \item {The disentangled falling component retains a relatively constant spectral index of $\alpha = -0.88 \pm 0.10$, which is consistent with nonthermal optically thin gyrosynchrotron radiation.} This $\alpha$ value corresponds to a power-law index of $\delta_r = 2.33$, which implies a hard electron energy distribution similar to the radio flares reported in T25.
    \item The flat component exhibits a flux density of $0.64 \pm 0.14$ mJy. While a flat {spectrum} is reminiscent of optically thin free-free radiation, the flux density is much higher than theoretical values {($S_\nu \leq 0.032$ mJy)} and implies mass-loss rates beyond the theoretical maximum of $\dot{M} \approx 2.4 \times 10^{-9}$ M$_\sun{}$ yr$^{-1}$.
    \item The flux density of the flat component is similar to values expected of optically thick gyroresonance radiation, given the temperatures and volume emission measures of simultaneous X-ray data. The consistent presence of this would require either multiple gyroresonance sources around the surface of AU Mic or near the poles, such that these sources are seen across the rotational period. Further, these regions would need to be compact enough to not absorb flaring emission (e.g., less than $0.1$ R$_*$ based on the estimated source sizes of small optically thick flares from T25) and be distributed in a way that contributes to a flat, unpolarized composite spectrum, even if specific sources exhibit circularly polarized emission.
    \item Using time-resolved VLA spectra, we estimate the electron kinetic energy rate of the gyrosynchrotron component to be between $10^{27}$ to $10^{30}$ erg s$^{-1}$ for field strengths from 3,000 to 100 G, respectively. This can be comparable to stellar flares over long-time scales. However, the lower energy levels and lack of flaring activity in multiwavelength observations may indicate that any magnetic reconnection events that accelerate these radio-emitting particles do not produce enough impulsive energy to create responses in the lower chromosphere.
\end{enumerate}

Future wide-band observations will be needed to determine simultaneous behavior and constrain spectral shape and frequency relations across the 1 to 100 GHz gyrosynchrotron range in both quiescent radiation and flaring events.
For quiescent observations, it is likely that modeling \citep[via][etc.]{Cranmer2013, Kuznetsov2021} will be needed to disentangle contributions from multiple sources and explore what range of spectral shapes and particle properties can recreate observations, including observed flux densities, polarizations, and spectral indices. Extended observations may also find lower levels of the flat component, indicating large-scale changes to the magnetic activity of AU Mic over time.
In such a case, simultaneous X-ray observations are also needed to determine if the emission remains consistent with a combined optically thick gyroresonance component.

\begin{acknowledgements}
This work was supported by NASA ADAP award program Number 80NSSC21K0632 and NASA XMM-Newton Guest Observer AO-17 Award 80NSSC19K0665. 
I.I.T.\ acknowledges support from the NSF Graduate Research Fellowship Program (GRFP).
Y.N.\ acknowledges funding from NASA ADAP 80NSSC21K0632, NASA TESS Cycle 6 80NSSC24K0493, NASA NICER Cycle 6 80NSSC24K1194, and HST GO 17464.

{We thank an anonymous reviewer for insightful comments that have led to improvements of this manuscript.}
We thank Drs.\ Rodrigo H.\ Hinojosa, Wei-Chun Jao, Jamie R.\ Lomax, James E.\ Neff, Leonardo A.\ Paredes, and Jack Soutter for their contributions during early parts of the 2018 AU Mic Campaign. I.I.T.\ thanks Dr.~Aline Vidotto for early discussion on mass loss rates in stellar flares {and Dr.~Kai Ikuta for supplying TESS starspot model fits for the rotational modulation of AU Mic.}
Y.N.\ also would like to acknowledge the the relevant discussions in the International Space Science Institute (ISSI)
Workshop ``Stellar Magnetism and its Impact on (Exo)Planets (\url{https://workshops.issibern.ch/stellar-magnetism/})" (held on June 2-6, 2025).

The National Radio Astronomy Observatory is a facility of the National Science Foundation operated under cooperative agreement by Associated Universities, Inc. 
The Australia Telescope Compact Array is part of the Australia Telescope National Facility (grid.421683.a) which is funded by the Australian Government for operation as a National Facility managed by CSIRO. We acknowledge the Gomeroi people as the traditional owners of the Observatory site. 
This work is based on observations obtained with XMM-Newton, an ESA science mission with instruments and contributions directly funded by ESA Member States and NASA.
\end{acknowledgements}

\bibliography{tristan2024}

@ARTICLE{Allred2015,
       author = {{Allred}, Joel C. and {Kowalski}, Adam F. and {Carlsson}, Mats},
        title = "{A Unified Computational Model for Solar and Stellar Flares}",
      journal = {\apj},
         year = 2015,
        month = aug,
       volume = {809},
       number = {1},
          eid = {104},
        pages = {104},
          doi = {10.1088/0004-637X/809/1/104},
archivePrefix = {arXiv},
       eprint = {1507.04375},
 primaryClass = {astro-ph.SR},
       adsurl = {https://ui.adsabs.harvard.edu/abs/2015ApJ...809..104A},
          url = {https://doi.org/10.1088/0004-637X/809/1/104}
}

@article{ Augereau2006,
	author = {{Augereau, J.-C.} and {Beust, H.}},
	title = {On the AU Microscopii debris disk - Density profiles, grain properties, and dust dynamics},
	DOI= "10.1051/0004-6361:20054250",
	url= "https://doi.org/10.1051/0004-6361:20054250",
	journal = {A\&A},
	year = 2006,
	volume = 455,
	number = 3,
	pages = "987-999",
}

@article{Bloot2024,
author = {{Bloot}, S. and {Callingham, J. R.} and {Vedantham, H. K.} and {Kavanagh, R. D.} and {Pope, B. J. S.} and {Climent, J. B.} and {Guirado, J. C.} and {Peña-Moñino, L.} and {Pérez-Torres, M.}},
title = {Phenomenology and periodicity of radio emission from the stellar system AU Microscopii★},
DOI= "10.1051/0004-6361/202348065",
url= "https://dx.doi.org/10.1051/0004-6361/202348065",
journal = {A\&A},
year = 2024,
volume = 682,
pages = "A170",
}

@article{CASA_2022,
doi = {10.1088/1538-3873/ac9642},
url = {https://dx.doi.org/10.1088/1538-3873/ac9642},
year = {2022},
month = {nov},
publisher = {The Astronomical Society of the Pacific},
volume = {134},
number = {1041},
pages = {114501},
author = {{CASA Team} and Ben Bean and Sanjay Bhatnagar and Sandra Castro and Jennifer Donovan Meyer and Bjorn Emonts and Enrique Garcia and Robert Garwood and Kumar Golap and Justo Gonzalez Villalba and Pamela Harris and Yohei Hayashi and Josh Hoskins and Mingyu Hsieh and Preshanth Jagannathan and Wataru Kawasaki and Aard Keimpema and Mark Kettenis and Jorge Lopez and Joshua Marvil and Joseph Masters and Andrew McNichols and David Mehringer and Renaud Miel and George Moellenbrock and Federico Montesino and Takeshi Nakazato and Juergen Ott and Dirk Petry and Martin Pokorny and Ryan Raba and Urvashi Rau and Darrell Schiebel and Neal Schweighart and Srikrishna Sekhar and Kazuhiko Shimada and Des Small and Jan-Willem Steeb and Kanako Sugimoto and Ville Suoranta and Takahiro Tsutsumi and Ilse M. van Bemmel and Marjolein Verkouter and Akeem Wells and Wei Xiong and Arpad Szomoru and Morgan Griffith and Brian Glendenning and Jeff Kern},
title = {CASA, the Common Astronomy Software Applications for Radio Astronomy},
journal = {Publications of the Astronomical Society of the Pacific}
}

@InProceedings{Cox1985,
author="Cox, J. J.
and Gibson, D. M.",
editor="Hjellming, Robert M.
and Gibson, David M.",
title="Thermal Emission and Possible Rotational Modulation in AU Mic",
booktitle="Radio Stars",
year="1985",
publisher="Springer Netherlands",
address="Dordrecht",
pages="233--236",
isbn="978-94-009-5420-5",
url="https://dx.doi.org/10.1007/978-94-009-5420-5_32"
}

@ARTICLE{Dulk1985,
   author = {{Dulk}, G.~A.},
    title = "{Radio emission from the sun and stars}",
  journal = {\araa},
     year = 1985,
   volume = 23,
    pages = {169-224},
      doi = {10.1146/annurev.aa.23.090185.001125},
   adsurl = {http://adsabs.harvard.edu/abs/1985ARA%26A..23..169D},
      url = {https://dx.doi.org/10.1146/annurev.aa.23.090185.001125}
}

@article{GaiaDR2,
	author = {{Gaia Collaboration} and {Brown, A. G. A.} and {Vallenari, A.} and {Prusti, T.} and {de Bruijne, J. H. J.} and {Babusiaux, C.} and {Bailer-Jones, C. A. L.} and {Biermann, M.} and {Evans, D. W.} and {Eyer, L.} and {Jansen, F.} and {Jordi, C.} and {Klioner, S. A.} and {Lammers, U.} and {Lindegren, L.} and {Luri, X.} and {Mignard, F.} and {Panem, C.} and {Pourbaix, D.} and {Randich, S.} and {Sartoretti, P.} and {Siddiqui, H. I.} and {Soubiran, C.} and {van Leeuwen, F.} and {Walton, N. A.} and {Arenou, F.} and {Bastian, U.} and {Cropper, M.} and {Drimmel, R.} and {Katz, D.} and {Lattanzi, M. G.} and {Bakker, J.} and {Cacciari, C.} and {Castañeda, J.} and {Chaoul, L.} and {Cheek, N.} and {De Angeli, F.} and {Fabricius, C.} and {Guerra, R.} and {Holl, B.} and {Masana, E.} and {Messineo, R.} and {Mowlavi, N.} and {Nienartowicz, K.} and {Panuzzo, P.} and {Portell, J.} and {Riello, M.} and {Seabroke, G. M.} and {Tanga, P.} and {Thévenin, F.} and {Gracia-Abril, G.} and {Comoretto, G.} and {Garcia-Reinaldos, M.} and {Teyssier, D.} and {Altmann, M.} and {Andrae, R.} and {Audard, M.} and {Bellas-Velidis, I.} and {Benson, K.} and {Berthier, J.} and {Blomme, R.} and {Burgess, P.} and {Busso, G.} and {Carry, B.} and {Cellino, A.} and {Clementini, G.} and {Clotet, M.} and {Creevey, O.} and {Davidson, M.} and {De Ridder, J.} and {Delchambre, L.} and {Dell’Oro, A.} and {Ducourant, C.} and {Fernández-Hernández, J.} and {Fouesneau, M.} and {Frémat, Y.} and {Galluccio, L.} and {García-Torres, M.} and {González-Núñez, J.} and {González-Vidal, J. J.} and {Gosset, E.} and {Guy, L. P.} and {Halbwachs, J.-L.} and {Hambly, N. C.} and {Harrison, D. L.} and {Hernández, J.} and {Hestroffer, D.} and {Hodgkin, S. T.} and {Hutton, A.} and {Jasniewicz, G.} and {Jean-Antoine-Piccolo, A.} and {Jordan, S.} and {Korn, A. J.} and {Krone-Martins, A.} and {Lanzafame, A. C.} and {Lebzelter, T.} and {Löffler, W.} and {Manteiga, M.} and {Marrese, P. M.} and {Martín-Fleitas, J. M.} and {Moitinho, A.} and {Mora, A.} and {Muinonen, K.} and {Osinde, J.} and {Pancino, E.} and {Pauwels, T.} and {Petit, J.-M.} and {Recio-Blanco, A.} and {Richards, P. J.} and {Rimoldini, L.} and {Robin, A. C.} and {Sarro, L. M.} and {Siopis, C.} and {Smith, M.} and {Sozzetti, A.} and {Süveges, M.} and {Torra, J.} and {van Reeven, W.} and {Abbas, U.} and {Abreu Aramburu, A.} and {Accart, S.} and {Aerts, C.} and {Altavilla, G.} and {Álvarez, M. A.} and {Alvarez, R.} and {Alves, J.} and {Anderson, R. I.} and {Andrei, A. H.} and {Anglada Varela, E.} and {Antiche, E.} and {Antoja, T.} and {Arcay, B.} and {Astraatmadja, T. L.} and {Bach, N.} and {Baker, S. G.} and {Balaguer-Núñez, L.} and {Balm, P.} and {Barache, C.} and {Barata, C.} and {Barbato, D.} and {Barblan, F.} and {Barklem, P. S.} and {Barrado, D.} and {Barros, M.} and {Barstow, M. A.} and {Bartholomé Muñoz, S.} and {Bassilana, J.-L.} and {Becciani, U.} and {Bellazzini, M.} and {Berihuete, A.} and {Bertone, S.} and {Bianchi, L.} and {Bienaymé, O.} and {Blanco-Cuaresma, S.} and {Boch, T.} and {Boeche, C.} and {Bombrun, A.} and {Borrachero, R.} and {Bossini, D.} and {Bouquillon, S.} and {Bourda, G.} and {Bragaglia, A.} and {Bramante, L.} and {Breddels, M. A.} and {Bressan, A.} and {Brouillet, N.} and {Brüsemeister, T.} and {Brugaletta, E.} and {Bucciarelli, B.} and {Burlacu, A.} and {Busonero, D.} and {Butkevich, A. G.} and {Buzzi, R.} and {Caffau, E.} and {Cancelliere, R.} and {Cannizzaro, G.} and {Cantat-Gaudin, T.} and {Carballo, R.} and {Carlucci, T.} and {Carrasco, J. M.} and {Casamiquela, L.} and {Castellani, M.} and {Castro-Ginard, A.} and {Charlot, P.} and {Chemin, L.} and {Chiavassa, A.} and {Cocozza, G.} and {Costigan, G.} and {Cowell, S.} and {Crifo, F.} and {Crosta, M.} and {Crowley, C.} and {Cuypers†, J.} and {Dafonte, C.} and {Damerdji, Y.} and {Dapergolas, A.} and {David, P.} and {David, M.} and {de Laverny, P.} and {De Luise, F.} and {De March, R.} and {de Martino, D.} and {de Souza, R.} and {de Torres, A.} and {Debosscher, J.} and {del Pozo, E.} and {Delbo, M.} and {Delgado, A.} and {Delgado, H. E.} and {Di Matteo, P.} and {Diakite, S.} and {Diener, C.} and {Distefano, E.} and {Dolding, C.} and {Drazinos, P.} and {Durán, J.} and {Edvardsson, B.} and {Enke, H.} and {Eriksson, K.} and {Esquej, P.} and {Eynard Bontemps, G.} and {Fabre, C.} and {Fabrizio, M.} and {Faigler, S.} and {Falcão, A. J.} and {Farràs Casas, M.} and {Federici, L.} and {Fedorets, G.} and {Fernique, P.} and {Figueras, F.} and {Filippi, F.} and {Findeisen, K.} and {Fonti, A.} and {Fraile, E.} and {Fraser, M.} and {Frézouls, B.} and {Gai, M.} and {Galleti, S.} and {Garabato, D.} and {García-Sedano, F.} and {Garofalo, A.} and {Garralda, N.} and {Gavel, A.} and {Gavras, P.} and {Gerssen, J.} and {Geyer, R.} and {Giacobbe, P.} and {Gilmore, G.} and {Girona, S.} and {Giuffrida, G.} and {Glass, F.} and {Gomes, M.} and {Granvik, M.} and {Gueguen, A.} and {Guerrier, A.} and {Guiraud, J.} and {Gutiérrez-Sánchez, R.} and {Haigron, R.} and {Hatzidimitriou, D.} and {Hauser, M.} and {Haywood, M.} and {Heiter, U.} and {Helmi, A.} and {Heu, J.} and {Hilger, T.} and {Hobbs, D.} and {Hofmann, W.} and {Holland, G.} and {Huckle, H. E.} and {Hypki, A.} and {Icardi, V.} and {Janßen, K.} and {Jevardat de Fombelle, G.} and {Jonker, P. G.} and {Juhász, Á. L.} and {Julbe, F.} and {Karampelas, A.} and {Kewley, A.} and {Klar, J.} and {Kochoska, A.} and {Kohley, R.} and {Kolenberg, K.} and {Kontizas, M.} and {Kontizas, E.} and {Koposov, S. E.} and {Kordopatis, G.} and {Kostrzewa-Rutkowska, Z.} and {Koubsky, P.} and {Lambert, S.} and {Lanza, A. F.} and {Lasne, Y.} and {Lavigne, J.-B.} and {Le Fustec, Y.} and {Le Poncin-Lafitte, C.} and {Lebreton, Y.} and {Leccia, S.} and {Leclerc, N.} and {Lecoeur-Taibi, I.} and {Lenhardt, H.} and {Leroux, F.} and {Liao, S.} and {Licata, E.} and {Lindstrøm, H. E. P.} and {Lister, T. A.} and {Livanou, E.} and {Lobel, A.} and {López, M.} and {Managau, S.} and {Mann, R. G.} and {Mantelet, G.} and {Marchal, O.} and {Marchant, J. M.} and {Marconi, M.} and {Marinoni, S.} and {Marschalkó, G.} and {Marshall, D. J.} and {Martino, M.} and {Marton, G.} and {Mary, N.} and {Massari, D.} and {Matijevič, G.} and {Mazeh, T.} and {McMillan, P. J.} and {Messina, S.} and {Michalik, D.} and {Millar, N. R.} and {Molina, D.} and {Molinaro, R.} and {Molnár, L.} and {Montegriffo, P.} and {Mor, R.} and {Morbidelli, R.} and {Morel, T.} and {Morris, D.} and {Mulone, A. F.} and {Muraveva, T.} and {Musella, I.} and {Nelemans, G.} and {Nicastro, L.} and {Noval, L.} and {O’Mullane, W.} and {Ordénovic, C.} and {Ordóñez-Blanco, D.} and {Osborne, P.} and {Pagani, C.} and {Pagano, I.} and {Pailler, F.} and {Palacin, H.} and {Palaversa, L.} and {Panahi, A.} and {Pawlak, M.} and {Piersimoni, A. M.} and {Pineau, F.-X.} and {Plachy, E.} and {Plum, G.} and {Poggio, E.} and {Poujoulet, E.} and {Prša, A.} and {Pulone, L.} and {Racero, E.} and {Ragaini, S.} and {Rambaux, N.} and {Ramos-Lerate, M.} and {Regibo, S.} and {Reylé, C.} and {Riclet, F.} and {Ripepi, V.} and {Riva, A.} and {Rivard, A.} and {Rixon, G.} and {Roegiers, T.} and {Roelens, M.} and {Romero-Gómez, M.} and {Rowell, N.} and {Royer, F.} and {Ruiz-Dern, L.} and {Sadowski, G.} and {Sagristà Sellés, T.} and {Sahlmann, J.} and {Salgado, J.} and {Salguero, E.} and {Sanna, N.} and {Santana-Ros, T.} and {Sarasso, M.} and {Savietto, H.} and {Schultheis, M.} and {Sciacca, E.} and {Segol, M.} and {Segovia, J. C.} and {Ségransan, D.} and {Shih, I-C.} and {Siltala, L.} and {Silva, A. F.} and {Smart, R. L.} and {Smith, K. W.} and {Solano, E.} and {Solitro, F.} and {Sordo, R.} and {Soria Nieto, S.} and {Souchay, J.} and {Spagna, A.} and {Spoto, F.} and {Stampa, U.} and {Steele, I. A.} and {Steidelmüller, H.} and {Stephenson, C. A.} and {Stoev, H.} and {Suess, F. F.} and {Surdej, J.} and {Szabados, L.} and {Szegedi-Elek, E.} and {Tapiador, D.} and {Taris, F.} and {Tauran, G.} and {Taylor, M. B.} and {Teixeira, R.} and {Terrett, D.} and {Teyssandier, P.} and {Thuillot, W.} and {Titarenko, A.} and {Torra Clotet, F.} and {Turon, C.} and {Ulla, A.} and {Utrilla, E.} and {Uzzi, S.} and {Vaillant, M.} and {Valentini, G.} and {Valette, V.} and {van Elteren, A.} and {Van Hemelryck, E.} and {van Leeuwen, M.} and {Vaschetto, M.} and {Vecchiato, A.} and {Veljanoski, J.} and {Viala, Y.} and {Vicente, D.} and {Vogt, S.} and {von Essen, C.} and {Voss, H.} and {Votruba, V.} and {Voutsinas, S.} and {Walmsley, G.} and {Weiler, M.} and {Wertz, O.} and {Wevers, T.} and {Wyrzykowski, Ł.} and {Yoldas, A.} and {Žerjal, M.} and {Ziaeepour, H.} and {Zorec, J.} and {Zschocke, S.} and {Zucker, S.} and {Zurbach, C.} and {Zwitter, T.}},
	title = {Gaia Data Release 2 - Summary of the contents and survey properties},
	DOI= "10.1051/0004-6361/201833051",
	url= "https://doi.org/10.1051/0004-6361/201833051",
	journal = {A\&A},
	year = 2018,
	volume = 616,
	pages = "A1",
}

@ARTICLE{Gudel1989,
       author = {{Güdel}, M. and {Benz}, A.~O.},
        title = "{Broad-band spectrum of dMe star radio emission.}",
      journal = {\aap},
         year = 1989,
        month = feb,
       volume = {211},
        pages = {L5-L8},
       url = {https://ui.adsabs.harvard.edu/abs/1989A&A...211L...5G}
}

@article{Kowalski2017A,
doi = {10.3847/1538-4357/836/1/12},
url = {https://dx.doi.org/10.3847/1538-4357/836/1/12},
year = {2017},
month = {feb},
publisher = {The American Astronomical Society},
volume = {836},
number = {1},
pages = {12},
author = {Adam F. Kowalski and Joel C. Allred and Adrian Daw and Gianna Cauzzi and Mats Carlsson},
title = {The Atmospheric Response to High Nonthermal Electron Beam Fluxes in Solar Flares. I. Modeling the Brightest NUV Footpoints in the X1 Solar Flare of 2014 March 29},
journal = {The Astrophysical Journal}
}

@ARTICLE{Kowalski2022,
AUTHOR={Kowalski, Adam F. },
TITLE={Near-ultraviolet continuum modeling of the 1985 April 12 great flare of AD Leo},
JOURNAL={Frontiers in Astronomy and Space Sciences},
VOLUME={9},
YEAR={2022},
URL={https://www.frontiersin.org/journals/astronomy-and-space-sciences/articles/10.3389/fspas.2022.1034458},
DOI={10.3389/fspas.2022.1034458},
ISSN={2296-987X}}

@ARTICLE{Kowalski2024b,
       author = {{Kowalski}, Adam F. and {Osten}, Rachel A. and {Notsu}, Yuta and {Tristan}, Isaiah I. and {Segura}, Antigona and {Maehara}, Hiroyuki and {Namekata}, Kosuke and {Inoue}, Shun},
        title = "{Rising Near-Ultraviolet Spectra in Stellar Megaflares}",
      journal = {arXiv e-prints},
         year = {2024b},
        month = nov,
          eid = {arXiv:2411.07913},
        pages = {arXiv:2411.07913},
          doi = {10.48550/arXiv.2411.07913},
archivePrefix = {arXiv},
       eprint = {2411.07913},
 primaryClass = {astro-ph.SR},
       adsurl = {https://ui.adsabs.harvard.edu/abs/2024arXiv241107913K},
          url = {https://dx.doi.org/10.48550/arXiv.2411.07913}
}

@ARTICLE{Kowalski2024c,
       author = {{Kowalski}, Adam F.},
        title = "{Stellar flares}",
      journal = {Living Reviews in Solar Physics},
         year = {2024c},
        month = dec,
       volume = {21},
       number = {1},
          eid = {1},
        pages = {1},
          doi = {10.1007/s41116-024-00039-4},
archivePrefix = {arXiv},
       eprint = {2402.07885},
 primaryClass = {astro-ph.SR},
       adsurl = {https://ui.adsabs.harvard.edu/abs/2024LRSP...21....1K},
          url = {https://dx.doi.org/10.1007/s41116-024-00039-4}
}

@article{Kuznetsov2021,
doi = {10.3847/1538-4357/ac29c0},
url = {https://dx.doi.org/10.3847/1538-4357/ac29c0},
year = {2021},
month = {nov},
publisher = {The American Astronomical Society},
volume = {922},
number = {2},
pages = {103},
author = {Alexey A. Kuznetsov and Gregory D. Fleishman},
title = {Ultimate Fast Gyrosynchrotron Codes},
journal = {The Astrophysical Journal}
}

@ARTICLE{Leto2000,
       author = {{Leto}, G. and {Pagano}, I. and {Linsky}, J.~L. and {Rodon{\`o}}, M. and {Umana}, G.},
        title = "{VLA observation of dMe stars}",
      journal = {\aap},
         year = 2000,
        month = jul,
       volume = {359},
        pages = {1035-1041},
          url = {https://ui.adsabs.harvard.edu/abs/2000A&A...359.1035L}
}

@article{MacGregor2020,
doi = {10.3847/1538-4357/ab711d},
url = {https://dx.doi.org/10.3847/1538-4357/ab711d},
year = {2020},
month = {mar},
publisher = {The American Astronomical Society},
volume = {891},
number = {1},
pages = {80},
author = {A. Meredith MacGregor and Rachel A. Osten and A. Meredith Hughes},
title = {Properties of M Dwarf Flares at Millimeter Wavelengths},
journal = {The Astrophysical Journal}
}

@article{Nita2002,
doi = {10.1086/339577},
url = {https://dx.doi.org/10.1086/339577},
year = {2002},
month = {may},
publisher = {},
volume = {570},
number = {1},
pages = {423},
author = {Gelu M. Nita and Dale E. Gary and L. J. Lanzerotti and D. J. Thomson},
title = {The Peak Flux Distribution of Solar Radio Bursts},
journal = {The Astrophysical Journal}
}

@article{Nita2004,
doi = {10.1086/382219},
url = {https://dx.doi.org/10.1086/382219},
year = {2004},
month = {apr},
publisher = {},
volume = {605},
number = {1},
pages = {528},
author = {Gelu M. Nita and Dale E. Gary and Jeongwoo Lee},
title = {Statistical Study of Two Years of Solar Flare Radio Spectra Obtained with the Owens Valley Solar Array},
journal = {The Astrophysical Journal}
}

@article{Osten2004,
doi = {10.1086/420770},
url = {https://dx.doi.org/10.1086/420770},
year = {2004},
month = {jul},
publisher = {},
volume = {153},
number = {1},
pages = {317},
author = {Rachel A. Osten and Alexander Brown and Thomas R. Ayres and Stephen A. Drake and Elena Franciosini and Roberto Pallavicini and Gianpiero Tagliaferri and Ron T. Stewart and Stephen L. Skinner and Jeffrey L. Linsky},
title = {A Multiwavelength Perspective of Flares on HR 1099: 4 Years of Coordinated Campaigns},
journal = {The Astrophysical Journal Supplement Series}
}

@article{Osten2005,
doi = {10.1086/427275},
url = {https://dx.doi.org/10.1086/427275},
year = {2005},
month = {mar},
publisher = {},
volume = {621},
number = {1},
pages = {398},
author = {Rachel A. Osten and Suzanne L. Hawley and Joel C. Allred and Christopher M. Johns-Krull and Christine Roark},
title = {From Radio to X-Ray: Flares on the dMe Flare Star EV Lacertae},
journal = {The Astrophysical Journal}
}

@article{Osten2006,
doi = {10.1086/504889},
url = {https://dx.doi.org/10.1086/504889},
year = {2006},
month = {aug},
publisher = {},
volume = {647},
number = {2},
pages = {1349},
author = {Osten, Rachel A. and Hawley, Suzanne L. and Allred, Joel and Johns-Krull, Christopher M. and Brown, Alexander and Harper, Graham M.},
title = {From Radio to X-Ray: The Quiescent Atmosphere of the dMe Flare Star EV Lacertae},
journal = {The Astrophysical Journal}
}

@ARTICLE{Plavchan2020,
       author = {{Plavchan}, Peter and {Barclay}, Thomas and {Gagn{\'e}}, Jonathan and {Gao}, Peter and {Cale}, Bryson and {Matzko}, William and {Dragomir}, Diana and {Quinn}, Sam and {Feliz}, Dax and {Stassun}, Keivan and {Crossfield}, Ian J.~M. and {Berardo}, David A. and {Latham}, David W. and {Tieu}, Ben and {Anglada-Escud{\'e}}, Guillem and {Ricker}, George and {Vanderspek}, Roland and {Seager}, Sara and {Winn}, Joshua N. and {Jenkins}, Jon M. and {Rinehart}, Stephen and {Krishnamurthy}, Akshata and {Dynes}, Scott and {Doty}, John and {Adams}, Fred and {Afanasev}, Dennis A. and {Beichman}, Chas and {Bottom}, Mike and {Bowler}, Brendan P. and {Brinkworth}, Carolyn and {Brown}, Carolyn J. and {Cancino}, Andrew and {Ciardi}, David R. and {Clampin}, Mark and {Clark}, Jake T. and {Collins}, Karen and {Davison}, Cassy and {Foreman-Mackey}, Daniel and {Furlan}, Elise and {Gaidos}, Eric J. and {Geneser}, Claire and {Giddens}, Frank and {Gilbert}, Emily and {Hall}, Ryan and {Hellier}, Coel and {Henry}, Todd and {Horner}, Jonathan and {Howard}, Andrew W. and {Huang}, Chelsea and {Huber}, Joseph and {Kane}, Stephen R. and {Kenworthy}, Matthew and {Kielkopf}, John and {Kipping}, David and {Klenke}, Chris and {Kruse}, Ethan and {Latouf}, Natasha and {Lowrance}, Patrick and {Mennesson}, Bertrand and {Mengel}, Matthew and {Mills}, Sean M. and {Morton}, Tim and {Narita}, Norio and {Newton}, Elisabeth and {Nishimoto}, America and {Okumura}, Jack and {Palle}, Enric and {Pepper}, Joshua and {Quintana}, Elisa V. and {Roberge}, Aki and {Roccatagliata}, Veronica and {Schlieder}, Joshua E. and {Tanner}, Angelle and {Teske}, Johanna and {Tinney}, C.~G. and {Vanderburg}, Andrew and {von Braun}, Kaspar and {Walp}, Bernie and {Wang}, Jason and {Wang}, Sharon Xuesong and {Weigand}, Denise and {White}, Russel and {Wittenmyer}, Robert A. and {Wright}, Duncan J. and {Youngblood}, Allison and {Zhang}, Hui and {Zilberman}, Perri},
        title = "{A planet within the debris disk around the pre-main-sequence star AU Microscopii}",
      journal = {\nat},
         year = 2020,
        month = jun,
       volume = {582},
       number = {7813},
        pages = {497-500},
          doi = {10.1038/s41586-020-2400-z},
archivePrefix = {arXiv},
       eprint = {2006.13248},
 primaryClass = {astro-ph.EP},
       adsurl = {https://ui.adsabs.harvard.edu/abs/2020Natur.582..497P},
          url = {https://dx.doi.org/10.1038/s41586-020-2400-z} 
}

@article{Robinson2001,
doi = {10.1086/321379},
url = {https://dx.doi.org/10.1086/321379},
year = {2001},
month = {jun},
publisher = {},
volume = {554},
number = {1},
pages = {368},
author = {Richard D. Robinson and Jeffrey L. Linsky and Bruce E. Woodgate and John G. Timothy},
title = {Far-Ultraviolet Observations of Flares on the
dM0e Star AU Microscopii},
journal = {The Astrophysical Journal}
}

@INPROCEEDINGS{Sault1995,
       author = {{Sault}, R.~J. and {Teuben}, P.~J. and {Wright}, M.~C.~H.},
        title = "{A Retrospective View of MIRIAD}",
    booktitle = {Astronomical Data Analysis Software and Systems IV},
         year = 1995,
       editor = {{Shaw}, R.~A. and {Payne}, H.~E. and {Hayes}, J.~J.~E.},
       series = {Astronomical Society of the Pacific Conference Series},
       volume = {77},
        month = jan,
        pages = {433},
          doi = {10.48550/arXiv.astro-ph/0612759},
archivePrefix = {arXiv},
       eprint = {astro-ph/0612759},
 primaryClass = {astro-ph},
       adsurl = {https://ui.adsabs.harvard.edu/abs/1995ASPC...77..433S},
          url = {https://dx.doi.org/10.485505/arXiv.astro-ph/0612759}
}

@ARTICLE{Smith2005,
   author = {{Smith}, K. and {G{\"u}del}, M. and {Audard}, M.},
    title = "{Flares observed with XMM-Newton and the VLA}",
  journal = {\aap},
   eprint = {astro-ph/0503022},
     year = 2005,
    month = jun,
   volume = 436,
    pages = {241-251},
      doi = {10.1051/0004-6361:20042054},
   adsurl = {http://adsabs.harvard.edu/abs/2005A%26A...436..241S},
      url = {https://dx.doi.org/10.1051/0004-6361:20042054}
}

@article{Tristan2023,
doi = {10.3847/1538-4357/acc94f},
url = {https://dx.doi.org/10.3847/1538-4357/acc94f},
year = {2023},
month = {jun},
publisher = {The American Astronomical Society},
volume = {951},
number = {1},
pages = {33},
author = {Isaiah I. Tristan and Yuta Notsu and Adam F. Kowalski and Alexander Brown and John P. Wisniewski and Rachel A. Osten and Eliot H. Vrijmoet and Graeme L. White and Brad D. Carter and Carol A. Grady and Todd J. Henry and Rodrigo H. Hinojosa and Jamie R. Lomax and James E. Neff and Leonardo A. Paredes and Jack Soutter},
title = {A 7 Day Multiwavelength Flare Campaign on AU Mic. I. High-time-resolution Light Curves and the Thermal Empirical Neupert Effect},
journal = {The Astrophysical Journal}
}

@ARTICLE{White1994,
       author = {{White}, S.~M. and {Lim}, J. and {Kundu}, M.~R.},
        title = "{Radio Constraints on Coronal Models for dMe Stars}",
      journal = {\apj},
         year = 1994,
        month = feb,
       volume = {422},
        pages = {293},
          doi = {10.1086/173727},
       adsurl = {https://ui.adsabs.harvard.edu/abs/1994ApJ...422..293W},
          url = {https://dx.doi.org/10.1086/173727}
}

@ARTICLE{Tristan2025,
       author = {{Tristan}, Isaiah I. and {Osten}, Rachel A. and {Notsu}, Yuta and {Kowalski}, Adam F. and {Brown}, Alexander and {White}, Graeme L. and {Grady}, Carol A. and {Henry}, Todd J. and {Vrijmoet}, Eliot Halley},
        title = "{A 7 day Multiwavelength Flare Campaign on AU Mic. II. Electron Densities and Kinetic Energies from High-frequency Radio Flares}",
      journal = {\apj},
         year = 2025,
        month = jun,
       volume = {986},
       number = {1},
          eid = {53},
        pages = {53},
          doi = {10.3847/1538-4357/adc565},
archivePrefix = {arXiv},
       eprint = {2503.14624},
 primaryClass = {astro-ph.SR},
       adsurl = {https://ui.adsabs.harvard.edu/abs/2025ApJ...986...53T},
          url = {https://dx.doi.org/10.3847/1538-4357/adc565}
}

@ARTICLE{Lim1996,
       author = {{Lim}, Jeremy and {White}, Stephen M.},
        title = "{Limits to Mass Outflows from Late-Type Dwarf Stars}",
      journal = {\apjl},
         year = 1996,
        month = may,
       volume = {462},
        pages = {L91},
          doi = {10.1086/310038},
       adsurl = {https://ui.adsabs.harvard.edu/abs/1996ApJ...462L..91L},
          url = {https://dx.doi.org/10.1086/310038}
}

@article{Ikuta2023,
doi = {10.3847/1538-4357/acbd36},
url = {https://doi.org/10.3847/1538-4357/acbd36},
year = {2023},
month = {may},
publisher = {The American Astronomical Society},
volume = {948},
number = {1},
pages = {64},
author = {Ikuta, Kai and Namekata, Kosuke and Notsu, Yuta and Maehara, Hiroyuki and Okamoto, Soshi and Honda, Satoshi and Nogami, Daisaku and Shibata, Kazunari},
title = {Starspot Mapping with Adaptive Parallel Tempering. II. Application to TESS Data for M-dwarf Flare Stars AU Microscopii, YZ Canis Minoris, and EV Lacertae},
journal = {The Astrophysical Journal}
}

@ARTICLE{Cranmer2013,
       author = {{Cranmer}, Steven R. and {Wilner}, David J. and {MacGregor}, Meredith A.},
        title = "{Constraining a Model of Turbulent Coronal Heating for AU Microscopii with X-Ray, Radio, and Millimeter Observations}",
      journal = {\apj},
         year = 2013,
        month = aug,
       volume = {772},
       number = {2},
          eid = {149},
        pages = {149},
          doi = {10.1088/0004-637X/772/2/149},
archivePrefix = {arXiv},
       eprint = {1306.4567},
 primaryClass = {astro-ph.SR},
       adsurl = {https://ui.adsabs.harvard.edu/abs/2013ApJ...772..149C},
          url = {https://doi.org/10.1088/0004-637X/772/2/149}
}

@ARTICLE{MacGregor2013,
       author = {{MacGregor}, Meredith A. and {Wilner}, David J. and {Rosenfeld}, Katherine A. and {Andrews}, Sean M. and {Matthews}, Brenda and {Hughes}, A. Meredith and {Booth}, Mark and {Chiang}, Eugene and {Graham}, James R. and {Kalas}, Paul and {Kennedy}, Grant and {Sibthorpe}, Bruce},
        title = "{Millimeter Emission Structure in the First ALMA Image of the AU Mic Debris Disk}",
      journal = {\apjl},
         year = 2013,
        month = jan,
       volume = {762},
       number = {2},
          eid = {L21},
        pages = {L21},
          doi = {10.1088/2041-8205/762/2/L21},
archivePrefix = {arXiv},
       eprint = {1211.5148},
 primaryClass = {astro-ph.EP},
       adsurl = {https://ui.adsabs.harvard.edu/abs/2013ApJ...762L..21M},
          url = {https://doi.org/10.1088/2041-8205/762/2/L21}
}

@ARTICLE{Gary1981,
       author = {{Gary}, D.~E. and {Linsky}, J.~L.},
        title = "{First detection of nonflare microwave emission from the coronae of single late-type dwarf stars.}",
      journal = {\apj},
         year = 1981,
        month = nov,
       volume = {250},
        pages = {284-292},
          doi = {10.1086/159373},
       adsurl = {https://ui.adsabs.harvard.edu/abs/1981ApJ...250..284G},
          url = {https://doi.org/10.1086/159373}
}

@article{Ricker2015,
author = {George R. Ricker and Joshua N. Winn and Roland Vanderspek and David W. Latham and G{\'a}sp{\'a}r  {\'A}. Bakos and Jacob L. Bean and Zachory K. Berta-Thompson and Timothy M. Brown and Lars Buchhave and Nathaniel R. Butler and R. Paul Butler and William J. Chaplin and David B. Charbonneau and J{\o}rgen Christensen-Dalsgaard and Mark Clampin and Drake Deming and John P. Doty and Nathan De Lee and Courtney Dressing and Edward W. Dunham and Michael Endl and Fran{\c{c}}ois Fressin and Jian Ge and Thomas Henning and Matthew J. Holman and Andrew W. Howard and Shigeru Ida and Jon M. Jenkins and Garrett Jernigan and John Asher Johnson and Lisa Kaltenegger and Nobuyuki Kawai and Hans Kjeldsen and Gregory Laughlin and Alan M. Levine and Douglas Lin and Jack J. Lissauer and Phillip MacQueen and Geoffrey Marcy and Peter R. McCullough and Timothy D. Morton and Norio Narita and Martin Paegert and Enric Palle and Francesco Pepe and Joshua Pepper and Andreas Quirrenbach and Stephen A. Rinehart and Dimitar Sasselov and Bun’ei Sato and Sara Seager and Alessandro Sozzetti and Keivan G. Stassun and Peter Sullivan and Andrew Szentgyorgyi and Guillermo Torres and Stephane Udry and Joel Villasenor},
title = {{Transiting Exoplanet Survey Satellite}},
volume = {1},
journal = {Journal of Astronomical Telescopes, Instruments, and Systems},
number = {1},
publisher = {SPIE},
pages = {014003},
year = {2014},
doi = {10.1117/1.JATIS.1.1.014003},
URL = {https://doi.org/10.1117/1.JATIS.1.1.014003}
}

@article{Cranmer2011,
doi = {10.1088/0004-637X/741/1/54},
url = {https://dx.doi.org/10.1088/0004-637X/741/1/54},
year = {2011},
month = {oct},
publisher = {The American Astronomical Society},
volume = {741},
number = {1},
pages = {54},
author = {Cranmer, Steven R. and Saar, Steven H.},
title = {TESTING A PREDICTIVE THEORETICAL MODEL FOR THE MASS LOSS RATES OF COOL STARS},
journal = {The Astrophysical Journal}
}

@article{Burnham2004,
author = {Kenneth P. Burnham and David R. Anderson},
title ={Multimodel Inference: Understanding AIC and BIC in Model Selection},

journal = {Sociological Methods \& Research},
volume = {33},
number = {2},
pages = {261-304},
year = {2004},
doi = {10.1177/0049124104268644},

URL = { 
    
        https://doi.org/10.1177/0049124104268644
    
    

},
eprint = { 
    
        https://doi.org/10.1177/0049124104268644
    
    

}
}

@ARTICLE{Achterberg2001,
       author = {{Achterberg}, Abraham and {Gallant}, Yves A. and {Kirk}, John G. and {Guthmann}, Axel W.},
        title = "{Particle acceleration by ultrarelativistic shocks: theory and simulations}",
      journal = {\mnras},
         year = 2001,
        month = dec,
       volume = {328},
       number = {2},
        pages = {393-408},
          doi = {10.1046/j.1365-8711.2001.04851.x},
archivePrefix = {arXiv},
       eprint = {astro-ph/0107530},
 primaryClass = {astro-ph},
       adsurl = {https://ui.adsabs.harvard.edu/abs/2001MNRAS.328..393A}
}

@ARTICLE{Bell1978,
       author = {{Bell}, A.~R.},
        title = "{The acceleration of cosmic rays in shock fronts - I.}",
      journal = {\mnras},
         year = 1978,
        month = jan,
       volume = {182},
        pages = {147-156},
          doi = {10.1093/mnras/182.2.147},
       adsurl = {https://ui.adsabs.harvard.edu/abs/1978MNRAS.182..147B}
}

@ARTICLE{Guo2014,
       author = {{Guo}, Fan and {Li}, Hui and {Daughton}, William and {Liu}, Yi-Hsin},
        title = "{Formation of Hard Power Laws in the Energetic Particle Spectra Resulting from Relativistic Magnetic Reconnection}",
      journal = {\prl},
         year = 2014,
        month = oct,
       volume = {113},
       number = {15},
          eid = {155005},
        pages = {155005},
          doi = {10.1103/PhysRevLett.113.155005},
archivePrefix = {arXiv},
       eprint = {1405.4040},
 primaryClass = {astro-ph.HE},
       adsurl = {https://ui.adsabs.harvard.edu/abs/2014PhRvL.113o5005G}
}

@article{Wilson2011,
    author = {Wilson, Warwick E. and Ferris, R. H. and Axtens, P. and Brown, A. and Davis, E. and Hampson, G. and Leach, M. and Roberts, P. and Saunders, S. and Koribalski, B. S. and Caswell, J. L. and Lenc, E. and Stevens, J. and Voronkov, M. A. and Wieringa, M. H. and Brooks, K. and Edwards, P. G. and Ekers, R. D. and Emonts, B. and Hindson, L. and Johnston, S. and Maddison, S. T. and Mahony, E. K. and Malu, S. S. and Massardi, M. and Mao, M. Y. and McConnell, D. and Norris, R. P. and Schnitzeler, D. and Subrahmanyan, R. and Urquhart, J. S. and Thompson, M. A. and Wark, R. M.},
    title = {The Australia Telescope Compact Array Broad-band Backend: description and first results*},
    journal = {Monthly Notices of the Royal Astronomical Society},
    volume = {416},
    number = {2},
    pages = {832-856},
    year = {2011},
    month = {09},
    issn = {0035-8711},
    doi = {10.1111/j.1365-2966.2011.19054.x},
    url = {https://doi.org/10.1111/j.1365-2966.2011.19054.x},
    eprint = {https://academic.oup.com/mnras/article-pdf/416/2/832/18582593/mnras0416-0832.pdf},
}

@ARTICLE{Perley2011,
       author = {{Perley}, R.~A. and {Chandler}, C.~J. and {Butler}, B.~J. and {Wrobel}, J.~M.},
        title = "{The Expanded Very Large Array: A New Telescope for New Science}",
      journal = {\apjl},
         year = 2011,
        month = sep,
       volume = {739},
       number = {1},
          eid = {L1},
        pages = {L1},
          doi = {10.1088/2041-8205/739/1/L1},
archivePrefix = {arXiv},
       eprint = {1106.0532},
 primaryClass = {astro-ph.IM},
       adsurl = {https://ui.adsabs.harvard.edu/abs/2011ApJ...739L...1P},
          url = {https://doi.org/10.1088/2041-8205/739/1/L1}
}

@ARTICLE{Kochukhov2021,
       author = {{Kochukhov}, Oleg},
        title = "{Magnetic fields of M dwarfs}",
      journal = {\aapr},
         year = 2021,
        month = dec,
       volume = {29},
       number = {1},
          eid = {1},
        pages = {1},
          doi = {10.1007/s00159-020-00130-3},
archivePrefix = {arXiv},
       eprint = {2011.01781},
 primaryClass = {astro-ph.SR},
       adsurl = {https://ui.adsabs.harvard.edu/abs/2021A&ARv..29....1K},
   url = {https://doi.org/10.1007/s00159-020-00130-3}
}

@article{ Vidotto2013,
	author = {{Vidotto, A. A.} and {Jardine, M.} and {Morin, J.} and {Donati, J.-F.} and {Lang, P.} and {Russell, A. J. B.}},
	title = {Effects of M dwarf magnetic fields on potentially habitable
          planets},
	DOI= "10.1051/0004-6361/201321504",
	url= "https://doi.org/10.1051/0004-6361/201321504",
	journal = {A\&A},
	year = 2013,
	volume = 557,
	pages = "A67",
	month = "",
}

@article{Fichtinger2017,
	author = {{Fichtinger}, Bibiana and {Güdel, Manuel} and {Mutel, Robert L.} and {Hallinan, Gregg} and {Gaidos, Eric} and {Skinner, Stephen L.} and {Lynch, Christene} and {Gayley, Kenneth G.}},
	title = {Radio emission and mass loss rate limits   of four young solar-type stars},
	DOI= "10.1051/0004-6361/201629886",
	url= "https://doi.org/10.1051/0004-6361/201629886",
	journal = {A\&A},
	year = 2017,
	volume = 599,
	pages = "A127",
}

@ARTICLE{Gudel1993,
       author = {{Güdel}, Manuel and {Benz}, Arnold O.},
        title = "{X-Ray/Microwave Relation of Different Types of Active Stars}",
      journal = {\apjl},
         year = 1993,
        month = mar,
       volume = {405},
        pages = {L63},
          doi = {10.1086/186766},
       adsurl = {https://ui.adsabs.harvard.edu/abs/1993ApJ...405L..63G},
url = {https://doi.org/10.1086/186766}
}

@article{Bastian2018,
doi = {10.3847/1538-4357/aab3cb},
url = {https://doi.org/10.3847/1538-4357/aab3cb},
year = {2018},
month = {apr},
publisher = {The American Astronomical Society},
volume = {857},
number = {2},
pages = {133},
author = {Bastian, T. S. and Villadsen, J. and Maps, A. and Hallinan, G. and Beasley, A. J.},
title = {Radio Emission from the Exoplanetary System ϵ Eridani},
journal = {The Astrophysical Journal}
}

@ARTICLE{Plant2024,
       author = {{Plant}, Kathryn and {Hallinan}, Gregg and {Bastian}, Tim},
        title = "{Radio Emission from the Magnetically Active M Dwarf UV Ceti from 1 to 105 GHz}",
      journal = {\apj},
         year = 2024,
        month = jul,
       volume = {970},
       number = {1},
          eid = {56},
        pages = {56},
          doi = {10.3847/1538-4357/ad4356},
archivePrefix = {arXiv},
       eprint = {2406.17280},
 primaryClass = {astro-ph.SR},
       adsurl = {https://ui.adsabs.harvard.edu/abs/2024ApJ...970...56P}
}

@ARTICLE{Gudel1994,
       author = {{Güdel}, Manuel},
        title = "{Quiescent Microwave Emission from Late-Type Stars}",
      journal = {\apjs},
         year = 1994,
        month = feb,
       volume = {90},
        pages = {743},
          doi = {10.1086/191899},
       adsurl = {https://ui.adsabs.harvard.edu/abs/1994ApJS...90..743G},
url = {https://doi.org/10.1086/191899}
}

@ARTICLE{Henry2024,
       author = {{Henry}, Todd J. and {Jao}, Wei-Chun},
        title = "{The Character of M Dwarfs}",
      journal = {\araa},
         year = 2024,
        month = sep,
       volume = {62},
       number = {1},
        pages = {593-633},
          doi = {10.1146/annurev-astro-052722-102740},
       adsurl = {https://ui.adsabs.harvard.edu/abs/2024ARA&A..62..593H},
          url = {https://dx.doi.org/10.1146/annurev-astro-052722-102740}
}

@ARTICLE{Babcock1955,
       author = {{Babcock}, Horace W. and {Babcock}, Harold D.},
        title = "{The Sun's Magnetic Field, 1952-1954.}",
      journal = {\apj},
         year = 1955,
        month = mar,
       volume = {121},
        pages = {349},
          doi = {10.1086/145994},
       adsurl = {https://ui.adsabs.harvard.edu/abs/1955ApJ...121..349B},
      url = {https://dx.doi.org/10.1086/145994}
}

@ARTICLE{Notsu2025,
       author = {{Notsu}, Yuta and {Tristan}, Isaiah I. and {Osten}, Rachel A. and {Brown}, Alexander and {Kowalski}, Adam F. and {Grady}, Carol A.},
        title = "{A Seven-day Multiwavelength Flare Campaign on AU Mic. III. Quiescent and Flaring Properties of the X-Ray Spectra and Chromospheric Lines}",
      journal = {\apj},
         year = 2025,
        month = nov,
       volume = {993},
       number = {2},
          eid = {212},
        pages = {212},
          doi = {10.3847/1538-4357/ae0578},
       adsurl = {https://ui.adsabs.harvard.edu/abs/2025ApJ...993..212N}
}

@ARTICLE{Villadsen2014,
       author = {{Villadsen}, Jackie and {Hallinan}, Gregg and {Bourke}, Stephen and {G{\"u}del}, Manuel and {Rupen}, Michael},
        title = "{First Detection of Thermal Radio Emission from Solar-type Stars with the Karl G. Jansky Very Large Array}",
      journal = {\apj},
         year = 2014,
        month = jun,
       volume = {788},
       number = {2},
          eid = {112},
        pages = {112},
          doi = {10.1088/0004-637X/788/2/112},
archivePrefix = {arXiv},
       eprint = {1405.2341},
 primaryClass = {astro-ph.SR},
       adsurl = {https://ui.adsabs.harvard.edu/abs/2014ApJ...788..112V},
url = {https://dx.doi.org/10.1088/0004-637X/788/2/112}
}

@ARTICLE{Vidotto2021,
       author = {{Vidotto}, Aline A.},
        title = "{The evolution of the solar wind}",
      journal = {Living Reviews in Solar Physics},
         year = 2021,
        month = dec,
       volume = {18},
       number = {1},
          eid = {3},
        pages = {3},
          doi = {10.1007/s41116-021-00029-w},
archivePrefix = {arXiv},
       eprint = {2103.15748},
 primaryClass = {astro-ph.SR},
       adsurl = {https://ui.adsabs.harvard.edu/abs/2021LRSP...18....3V}
}

@ARTICLE{Strubbe2006,
       author = {{Strubbe}, Linda E. and {Chiang}, Eugene I.},
        title = "{Dust Dynamics, Surface Brightness Profiles, and Thermal Spectra of Debris Disks: The Case of AU Microscopii}",
      journal = {\apj},
         year = 2006,
        month = sep,
       volume = {648},
       number = {1},
        pages = {652-665},
          doi = {10.1086/505736},
archivePrefix = {arXiv},
       eprint = {astro-ph/0510527},
 primaryClass = {astro-ph},
       adsurl = {https://ui.adsabs.harvard.edu/abs/2006ApJ...648..652S}
}

@ARTICLE{Johnstone2015b,
       author = {{Johnstone}, C.~P. and {G{\"u}del}, M. and {Brott}, I. and {L{\"u}ftinger}, T.},
        title = "{Stellar winds on the main-sequence. II. The evolution of rotation and winds}",
      journal = {\aap},
         year = 2015,
        month = may,
       volume = {577},
          eid = {A28},
        pages = {A28},
          doi = {10.1051/0004-6361/201425301},
archivePrefix = {arXiv},
       eprint = {1503.07494},
 primaryClass = {astro-ph.SR},
       adsurl = {https://ui.adsabs.harvard.edu/abs/2015A&A...577A..28J}
}

@ARTICLE{Mukai1993,
       author = {{Mukai}, K.},
        title = "{PIMMS and Viewing: proposal preparation tools}",
      journal = {Legacy},
         year = 1993,
        month = may,
       volume = {3},
        pages = {21-31},
       adsurl = {https://ui.adsabs.harvard.edu/abs/1993Legac...3...21M}
}
\bibliographystyle{aasjournalv7}

\end{document}